\documentclass[fleqn,usenatbib,useAMS]{mnras}

\usepackage{graphicx}	
\usepackage{amsmath}	
\usepackage{amssymb}	
\usepackage{units}
\usepackage{gensymb}
\usepackage[flushleft]{threeparttable}
\newbox{\bigpicturebox}

\def\gridline#1{\vskip6pt\hbox to\hsize{#1}\vskip6pt}

\def\fig#1#2#3{\hfill\vbox{\parskip=0pt\hsize=#2
\includegraphics[width=#2]{#1}\vskip2pt\vtop{\centering
\footnotesize
\hsize=#2
#3\vskip1pt
}}\hfill}

\usepackage[T1]{fontenc}
\usepackage{ae,aecompl}

\usepackage{newtxtext,newtxmath}

\title[Binary-driven stellar rotation evolution]{Binary-driven stellar rotation evolution at the main-sequence turn-off in star clusters}

\author[W. Sun et al.]{
Weijia Sun$^{1,2,3,4}$\thanks{Contact e-mail: \href{mailto:this.is.weijia@gmail.com}{this.is.weijia@gmail.com}}
Richard de Grijs$^{3,4,5}$,
Licai Deng$^{2,6,7}$
and Michael D. Albrow$^{8}$
\\
$^{1}$Department of Astronomy, School of Physics, Peking University, Beijing 100871, China\\
$^{2}$Key Laboratory for Optical Astronomy, National Astronomical Observatories, Chinese Academy of Sciences, 20A Datun Road, Chaoyang District, Beijing 100012, China\\
$^{3}$Department of Physics and Astronomy, Macquarie University, Balaclava Road, Sydney, NSW 2109, Australia\\
$^{4}$Centre for Astronomy, Astrophysics and Astrophotonics, Macquarie University, Balaclava Road, Sydney, NSW 2109, Australia\\
$^{5}$International Space Science Institute--Beijing, 1 Nanertiao, Zhongguancun, Hai Dian District, Beijing 100190, China\\
$^{6}$School of Astronomy and Space Science, University of the Chinese Academy of Sciences, Huairou 101408, China\\
$^{7}$Department of Astronomy, China West Normal University, Nanchong 637002, China\\
$^{8}$School of Physical and Chemical Sciences, University of Canterbury, Private Bag 4800, Christchurch, New Zealand
}

\date{Accepted. Received; in original form}

\pubyear{2020}

\begin{document}
\label{firstpage}
\pagerange{\pageref{firstpage}--\pageref{lastpage}}
\maketitle

\begin{abstract}
The impact of stellar rotation on the morphology of star cluster
colour--magnitude diagrams is widely acknowledged. However, the
physics driving the distribution of the equatorial rotation velocities
of main-sequence turn-off (MSTO) stars is as yet poorly
understood. Using \textit{Gaia} Data Release 2 photometry and new
Southern African Large Telescope medium-resolution spectroscopy, we
analyse the intermediate-age ($\sim\unit[1]{Gyr}$-old) Galactic open
clusters NGC 3960, NGC 6134 and IC 4756 and develop a novel method to
derive their stellar rotation distributions based on SYCLIST stellar
rotation models. Combined with literature data for the open clusters
NGC 5822 and NGC 2818, we find a tight correlation between the number
ratio of slow rotators and the clusters' binary fractions. The
blue-main-sequence stars in at least two of our clusters are more
centrally concentrated than their red-main-sequence counterparts. The
origin of the equatorial stellar rotation distribution and its
evolution remains as yet unidentified. However, the observed
correlation in our open cluster sample suggests a binary-driven
formation mechanism.
\end{abstract}

\begin{keywords}
galaxies: star clusters: general -- techniques: spectroscopic
\end{keywords}



\begingroup
\let\clearpage\relax
\endgroup
\newpage

\section{Introduction}

Extended main-sequence turn-offs \citep[eMSTOs;
  e.g.,][]{2007MNRAS.379..151M, 2011ApJ...737....3G} and split main
sequences \citep[MSs; e.g.,][]{2015MNRAS.453.2637D,
  2017ApJ...844..119L} pose a fundamental challenge to our traditional
understanding of star clusters as `simple stellar populations'. Both
features are thought to be driven by differences in stellar rotation
rates \citep[e.g.,][]{2015MNRAS.453.2070N}, supported by an increasing
body of spectroscopic evidence in both Magellanic Cloud clusters
\citep{2017ApJ...846L...1D, 2020MNRAS.492.2177K} and Galactic open
clusters \citep[OCs;][]{2018MNRAS.480.3739B, 2019ApJ...876..113S,
  2019ApJ...883..182S}. Fast rotators appear redder than their slowly
rotating counterparts because of the compound effects of gravity
darkening and rotational mixing \citep{2013ApJ...776..112Y}.

As the effects of stellar rotation in defining the morphology of
cluster colour--magnitude diagrams (CMDs) are now well-understood, the
rotation distributions of MSTO stars and their origin in star clusters
represent the next key open question. A cluster mostly composed of
initially rapidly rotating stars may reproduce the observed
`converging' subgiant branch in NGC 419
\citep{2016ApJ...826L..14W}. \citet{2015MNRAS.453.2637D} argued that
split MSs are composed of two coeval populations with different
rotation rates: a slowly rotating blue-MS (bMS) and a rapidly rotating
red-MS (rMS) population. This bimodal rotational velocity distribution
was confirmed for NGC 2287 \citep{2019ApJ...883..182S}, where the
well-separated double MS in this young OC is tightly correlated with a
dichotomous distribution of stellar rotation
rates. \citet{2020MNRAS.492.2177K} also discovered a bimodal
distribution in the rotation rates of MSTO stars in the 1.5 Gyr-old
cluster NGC 1846. However, the peak rotational velocities in this
cluster are slower than those in the younger cluster NGC 2287, thus
offering a hint of stellar rotation evolution.

This naturally raises the question as to the origin of such a bimodal
rotation distribution. \citet{2015MNRAS.453.2637D,
  2017NatAs...1E.186D} suggested that bMS stars in young clusters
might have resulted from tidal braking of initially fast rotators on
time-scales of a few $\times 10^7$ yr. Alternatively, the bimodal
rotation distribution could have been established within the first few
Myr, regulated by either the disc-locking time-scale
\citep{2020MNRAS.495.1978B} or the accretion abundance of
circumstellar discs \citep{2020A&A...641A..73H}. In this paper, we
discover a tight correlation between the stellar rotation rates and
binary fractions in five intermediate-age Galactic OCs with similar
dynamical ages, favouring a binary-driven model. This is further
supported by the greater central concentration of the bMS stars
relative to the rMS population in at least two of our
clusters. Intriguingly, our observational results are inconsistent
with the prevailing theoretical models, which thus invites more
detailed future investigations.

This paper is organised as follows. We present our photometric and
spectroscopic data reduction in Section~\ref{sec:data}. Our analysis
to unravel the OCs' rotation distributions and its validation are
described in Section~\ref{sec:method}. A discussion about the
formation mechanism determining the stellar rotation rates in clusters
and our conclusions are summarised in Section~\ref{sec:discussion}.

\section{Observations and Data Reduction} \label{sec:data}
\subsection{Photometric data}

We obtained photometry and astrometry of cluster stars from
\textit{Gaia} Data Release (DR) 2 \citep{2016A&A...595A...2G,
  2018A&A...616A...1G} and determined their cluster membership
probabilities based on proper-motion and parallax analysis, as
follows.

First, we downloaded the \textit{Gaia} DR2 astrometry, proper motions,
photometry and parallaxes for all stars located within 1 degree of a
cluster's centre. We selected all sources with
$\texttt{parallax\_over\_error} > 1$ and a renormalised unit
weight
error\footnote{\url{https://www.cosmos.esa.int/web/gaia/dr2-known-issues}},
RUWE $< 1.4$. We also used the flux excess factor, $E = (I_\mathrm{BP}
+ I_\mathrm{RP})/I_\mathrm{G}$
(\texttt{phot\_bp\_rp\_excess\_factor})---where $I_\mathrm{X}$ is the
photometric flux in band $X$ \citep{2018A&A...616A...4E}---to exclude
possible issues with the \textit{Gaia} BP and RP photometry
\citep{2018A&A...616A..10G}:
\begin{equation}
1.0 + 0.015(G_\mathrm{BP} - G_\mathrm{RP})^2 < E < 1.3 + 0.06(G_\mathrm{BP} - G_\mathrm{RP})^2.
\end{equation}
The final step before member selection involved correcting the
\textit{Gaia} magnitudes for the effects of saturation at bright
magnitudes. This was done by employing the equations of
\citet{2018A&A...616A...4E}. We adopted the corrected magnitudes for
subsequent analysis.

\begin{figure*}
\gridline{
\fig{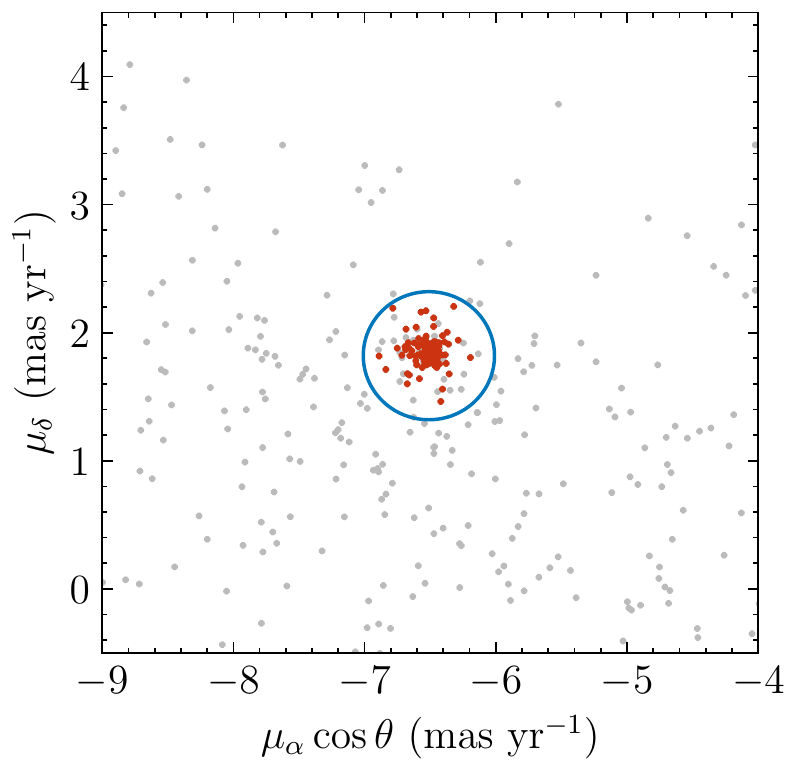}{0.3\textwidth}{}
\fig{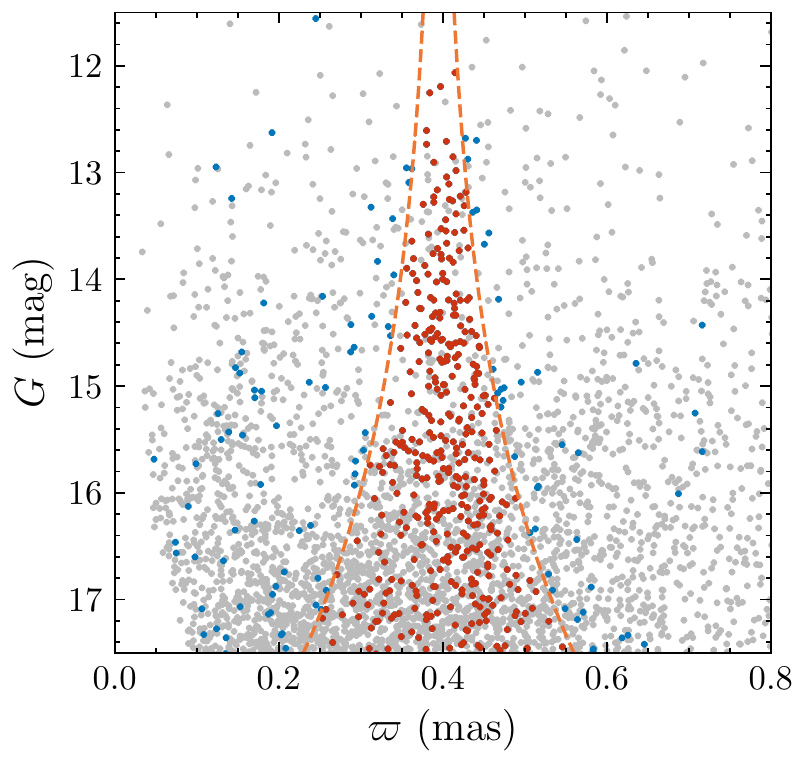}{0.3\textwidth}{}
\fig{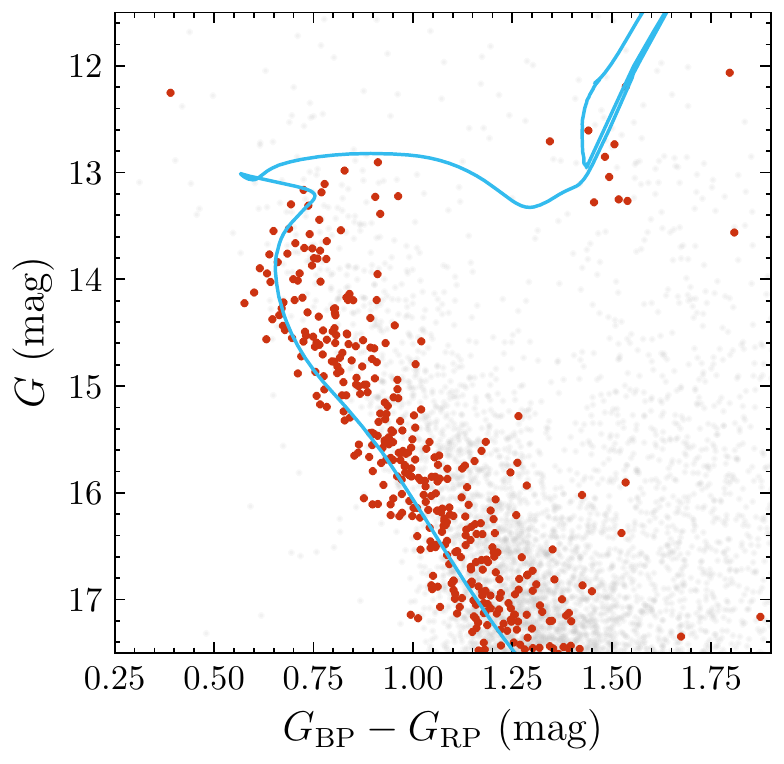}{0.3\textwidth}{}
}
\caption{Illustration of the procedure adopted to select probable
  members of NGC 3960. (left) VPD of the proper motions of bright
  field stars (grey) and bright cluster members (red) in the cluster's
  field. The primary selection boundary, based on proper motion
  constraints, is indicated by the blue circle
  ($\unit[1]{mas\,yr^{-1}}$). (middle) $G$-band photometry versus
  stellar parallaxes. Candidates from the primary selection step are
  shown as blue dots, whereas those selected based on their parallaxes
  are shown as red dots. The vertical dashed lines represent the
  parallax selection boundaries. (right) CMD of all stars in the field
  (grey) and the NGC 3960 member stars (red). The best-fitting
  isochrone to the bulk stellar population is also
  shown.\label{fig:member}}
\end{figure*}

Next, to derive a clean sample of member stars, we analysed the
vector-point diagram (VPD) of the stellar proper motions and located
the distribution's centre based on 2D kernel density estimation
(KDE). A cut in $\mu_\mathrm{R} = \sqrt{(\mu_\alpha\cos\theta-\langle
  \mu_\alpha\cos\theta \rangle)^2+(\mu_\delta-\langle \mu_\delta
  \rangle)^2}$ was applied to perform the primary membership
selection. We then further selected the remaining stars based on their
parallaxes, $\varpi$, rejecting all stars whose parallaxes deviated
from the mean value by more than four times the corresponding
r.m.s. (see Fig.~\ref{fig:member}). By comparing the CMDs composed of
our selection of cluster member stars with their literature
counterparts \citep{2018MNRAS.480.3739B, 2018ApJ...869..139C,
  2018A&A...615A..49C, 2019ApJ...876..113S}, we concluded that our
selection approach is indeed robust and reliable.

The clusters' ages, distances and extinction values were estimated
based on visual comparison with the MIST isochrones
\citep{2016ApJ...823..102C}. The parameters of the best-fitting
isochrone to the blue edge of the bulk stellar population of each
cluster (Fig.~\ref{fig:member}, right) are included in
Table~\ref{tab:oc}. Our independently derived cluster parameters are
consistent with literature values
\citep[e.g.,][]{2019A&A...623A.108B}; minor differences relate to the
choice of stellar models adopted.

\begin{table*}
\caption{Derived properties of our intermediate-age OCs\label{tab:oc}}
\begin{threeparttable}
\begin{tabular}{lccccccccc}
\hline
\hline
Cluster & $\log (t\,\unit{yr^{-1}})$ & $(m-M)_0$ & [Fe/H] & $A_V$ & $f_\mathrm{b}$ & $\sigma$ & $N_\mathrm{slow}/N_\mathrm{tot}$ & $\Lambda_\mathrm{MSR, b}$ & $\Lambda_\mathrm{MSR, r}$ \\
& & (\unit{mag}) & (\unit{dex}) & (\unit{mag}) & & (\unit{mag})\\
(1) & (2) & (3) & (4) & (5) & (6) & (7) & (8) & (9) & (10)\\
\hline
NGC 3960 & 8.98 & 11.35 & $-$0.25 & 1.00 & $0.26^{+0.05}_{-0.04}$ &
$0.06^{+0.003}_{-0.002}$ & $0.34^{+0.04}_{-0.04}$ & $2.74 \pm 0.25$ & $1.78 \pm 0.21$
\\[2pt]
NGC 6134 & 8.96 & 10.28 & 0.08 & 1.32 & $0.43^{+0.03}_{-0.03}$ &
$0.03^{+0.001}_{-0.002}$ & $0.55^{+0.05}_{-0.05}$ & $1.35 \pm 0.08$ &
--$^a$\\[2pt]
IC 4756 & 8.94 & 8.16 & 0.00 & 0.50 & $0.48^{+0.01}_{-0.02}$ &
$0.02^{+0.003}_{-0.004}$ & $0.74^{+0.03}_{-0.03}$ & $1.50 \pm 0.12$ &
--$^a$ \\[2pt]
NGC 5822 & 8.90 & 9.46 & 0.00 & 0.50 & $0.15^{+0.02}_{-0.02}$ &
$0.01^{+0.002}_{-0.002}$ & $0.24^{+0.03}_{-0.03}$ & $2.22 \pm 0.20$ & $1.28 \pm 0.12$
\\[2pt]
NGC 2818 & 8.89 & 12.56 & 0.00 & 0.60 & $0.28^{+0.03}_{-0.03}$ &
$0.02^{+0.005}_{-0.004}$ & $0.40^{+0.05}_{-0.06}$ & $1.19 \pm 0.11$ & $0.92 \pm 0.09$
\\[2pt]
\hline
\multicolumn{10}{l}{$^a$ Insufficient sample size.}
\end{tabular}
\begin{tablenotes}
\small
\item (1) Cluster name; (2) Age; (3) Distance modulus; (4)
  Metallicity; (5) Extinction; (6) Total binary fraction; (7) Scatter
  in pseudo-colour; (8) Slow-rotator number fraction among synthetic
  MSTO stars; (9, 10) Mass segregation ratios (MSRs) of (9) bMS and
  (10) rMS stars.
\end{tablenotes}
\end{threeparttable}
\end{table*}

\subsection{Spectroscopic data}

For three $\sim\unit[1]{Gyr}$-old Galactic OCs---NGC 3960, NGC 6134
and IC 4756---we obtained new medium-resolution spectroscopy. We
expanded our sample by including NGC 5822 \citep{2019ApJ...876..113S}
and NGC 2818 \citep{2018MNRAS.480.3739B}. The first four of these
clusters were observed in multi-object mode with the Robert Stobie
Spectrograph on the Southern African Large Telescope (programmes
2017-2-SCI-038, 2018-1-SCI-006 and 2018-2-SCI-002), during 18 nights
between 4 January 2018 and 26 April 2019.

The spectrograph configuration for our new observations was the same
as that adopted by \citet{2019ApJ...876..113S}. The PG2300 grating was
used at a grating angle of 34.25 degrees and a camera station angle of
68.5 degrees. This configuration, combined with a slit width of 1
arcsec, yields a central wavelength of \unit[4884.4]{\AA} and a
spectral coverage of $\sim$\unit[4345\textup{--}5373]{\AA} at a
resolving power of $R\sim 4000$ (the detector's wavelength coverage
also depends on the location of the slits). Argon lamp exposures and
flat-field calibration frames were taken at the end of each
observation for wavelength calibration and flat-field correction,
respectively. We did not observe standard-star spectra, since flux
calibration does not affect the line profiles of the Mg {\sc i}
triplets used for the rotational velocity estimation. Overscan
correction, bias subtraction, gain correction and wavelength
calibration were done using the PySALT package
\citep{2010SPIE.7737E..25C}. The exposure time was adjusted based on
the average brightness of the stars in the mask to ensure a typical
minimum signal-to-noise ratio (S/N) per pixel exceeding 200.

\begin{figure*}
\gridline{
\fig{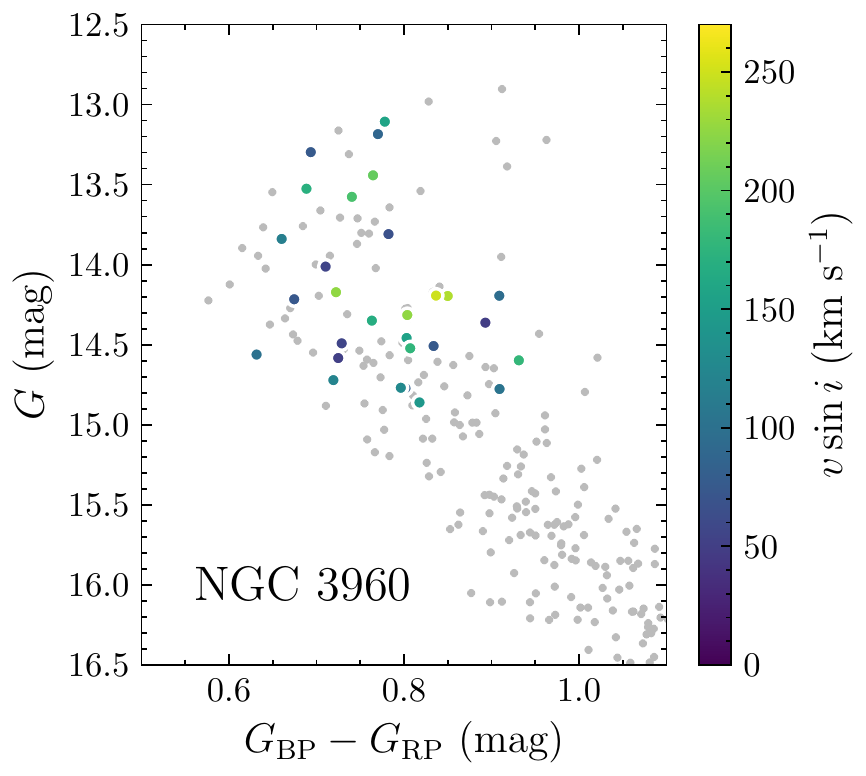}{0.3\textwidth}{}
\fig{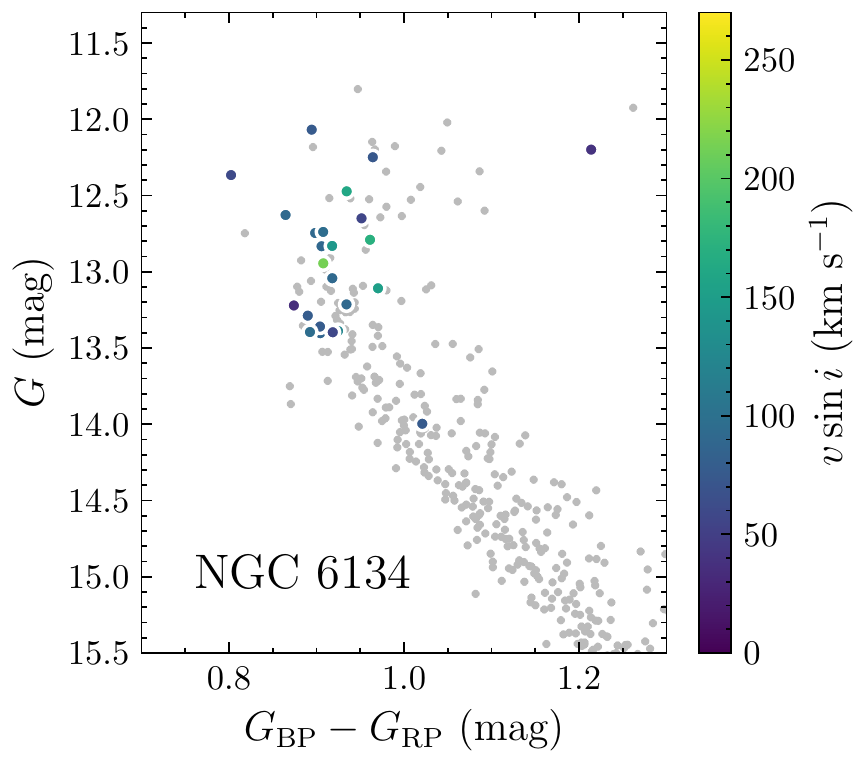}{0.3\textwidth}{}
\fig{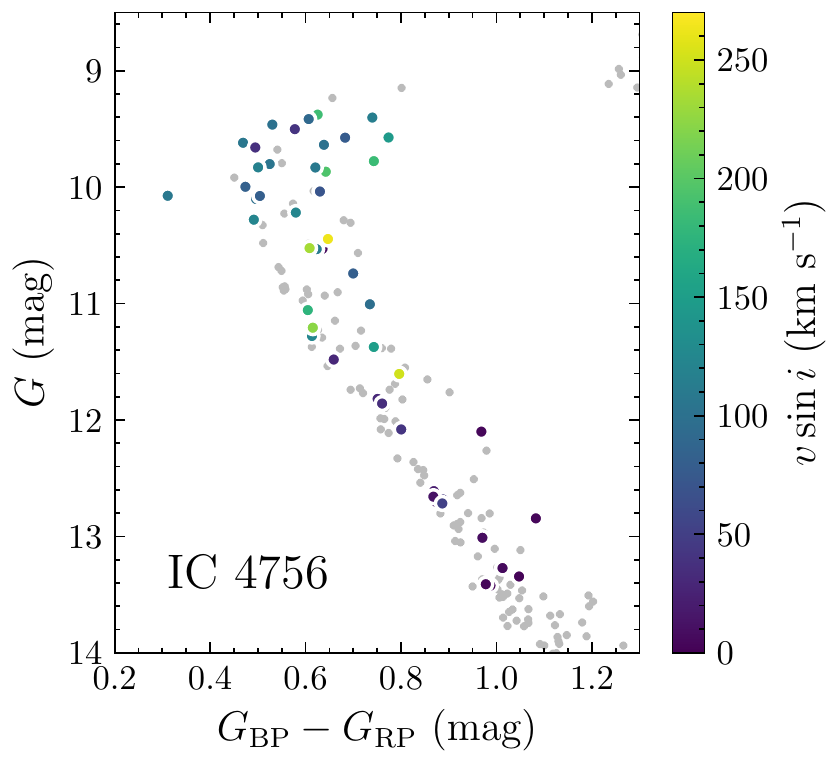}{0.3\textwidth}{}
}
\caption{CMDs of (left) NGC 3960, (middle) NGC 6134 and (right) IC
  4756 with their member stars colour-coded by their projected
  rotational velocities. Slow rotators (blue) are preferentially found
  on the blue side of the MSTO, whereas fast rotators (yellow) tend to
  be located on the red side. \label{fig:cmdrot}}
\end{figure*}

We determined the stars' projected rotational velocities following
\citet{2019ApJ...876..113S} by fitting \ion{H}{$\beta$} and the
\ion{Mg}{I} triplet with the synthetic stellar spectra from the Pollux
database \citep{2010A&A...516A..13P}, convolved with the rotational
profile for a given rotational velocity and implemented by adopting
instrumental broadening. The error was estimated through a comparison
of the rotational velocities of the mock data with those derived
through profile fitting \citep[][their figure
  4]{2019ApJ...876..113S}. For IC 4756 we also included stellar
rotation measurements from \citet{1984MNRAS.208...83S} and
\citet{2015A&A...580A..66S}. For stars with multi-epoch observations,
a variability test indicated no significant variation of either
  the radial or the rotational velocities in our data set (except for
  one star in NGC 6134; see Section~\ref{sec:discussion}). As such,
we adopted the average rotational velocity. We thus obtained rotation
measurements of 29, 25 and 49 stars in NGC 3960, NGC 6134 and IC 4756,
respectively. In Fig.~\ref{fig:cmdrot}, we present the CMDs of NGC
3960, NGC 6134 and IC 4756, with the member stars colour-coded by
their rotational velocities. Combined with 24 stars in NGC 5822 and 57
in NGC 2818, we achieved 20--30 per cent completeness across the eMSTO
for all five clusters. This is sufficient to reliably derive their
rotation distributions (see Section~\ref{sec:rot}).

\subsection{Colour--magnitude diagram}
\label{sec:cmd}

\begin{figure*}
\includegraphics{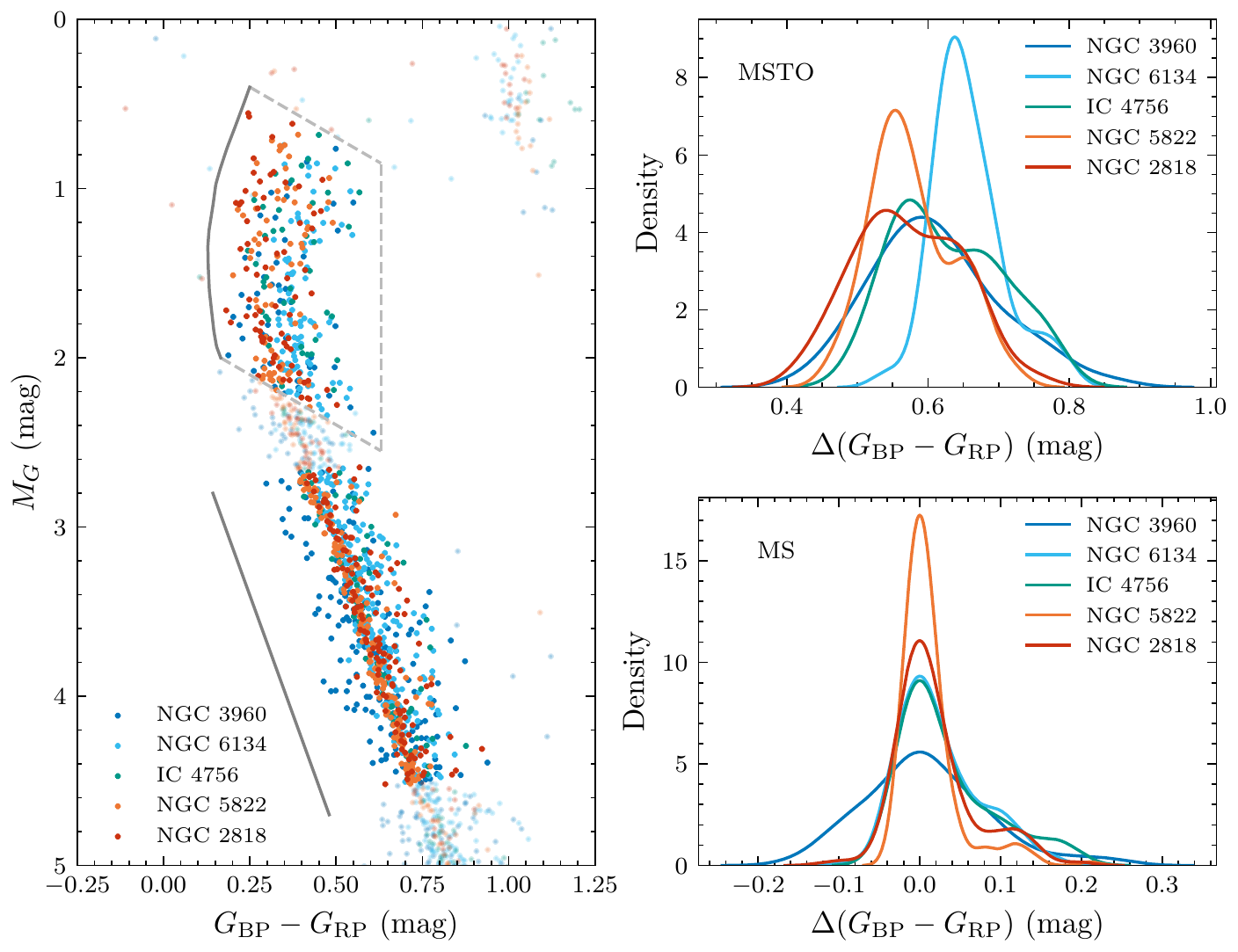}
\caption{(left) CMD of our cluster member stars in \textit{Gaia}
  passbands. The MSTO and MS samples are highlighted as solid
  dots. Grey dashed line: MSTO stars. Dark grey lines: Ridge lines
  used to calculate the populations' pseudo-colours. (right) KDEs of
  the (top) MSTO and (bottom) MS samples. The peaks of the MS
  distributions have been aligned and shifted for clarity and
  comparison. Any misalignment is owing to minor differences in the
  clusters' ages and metallicities. \label{fig:cmd}}
\end{figure*}

All sample clusters have similar chronological and dynamical ages (see
Appendix~\ref{sec:dynamical}), thus enabling direct comparison upon
correction of the clusters' photometry for extinction and distance
differences: see Fig.~\ref{fig:cmd} (left). Whereas the clusters'
MSTOs exhibit different patterns, their lower MSs converge at
approximately the same position in CMD space. We selected samples of
MSTO and MS stars to better illustrate their morphologies using
$\Delta(G_\mathrm{BP} - G_\mathrm{RP})$ pseudo-colour distributions.

The selection boundaries applied to the MSTO stars are shown in
Fig.~\ref{fig:cmd} (grey enclosure): stars bluer than $G_\mathrm{BP} -
G_\mathrm{RP} =$ \unit[0.63]{mag} and redder than the blue ridge line
defined by the rotation model (see Section~\ref{sec:method}) were
considered MSTO stars. The slopes of the cuts at the bright and faint
ends follow the expected stellar locus change caused by stellar
rotation. The $\Delta(G_\mathrm{BP} - G_\mathrm{RP})$ pseudo-colour is
defined as the difference in colour with respect to a cluster's ridge
line. We excluded a number of possible blue straggler stars at colours
bluer than the best-fitting isochrone to the cluster's bulk stellar
population minus $3\sigma$ (see Section~\ref{sec:method}). Next, we
defined a straight line parallel to the MS, at $2.8 \le M_G \le
\unit[4.5]{mag}$, as our MS reference. In Fig.~\ref{fig:cmd} (right),
KDEs of MSTO and MS stars are presented. The MS KDEs exhibit similar
profiles, with a dominant peak at $\unit[0]{mag}$ and a minor bump
corresponding to the binary sequence $\sim\unit[0.1]{mag}$ redder than
the peak. However, the MSTO KDEs exhibit significant variation both in
the peak locations and the overall distributions.

\section{Unravelling the rotation distributions} \label{sec:method}

Several recent attempts have been made to unravel cluster rotation
distributions \citep{2019ApJ...887..199G,
  2020MNRAS.491.2129D}. However, most of these estimates were based
solely on their CMD morphologies \citep[see
  also][]{2020MNRAS.492.2177K}, i.e., based on the probability of
models matching the observations using Hess diagrams. Whereas this may
be suitable for MC clusters, where the numbers of member stars are
sufficiently large and where it is difficult or impossible to obtain
direct velocity information, such analyses of OCs inevitably suffer
from stochastic sampling effects (see
Section~\ref{sec:rot}). Spectroscopic observations can ameliorate
these effects.

We used the SYCLIST stellar rotation models
\citep{2013A&A...553A..24G, 2014A&A...566A..21G} to generate synthetic
clusters, assuming solar metallicity ($Z = 0.014$), an age of $\log( t
\mbox{ yr}^{-1}) = 8.95$ and a 50 per cent binary fraction. The model
also considers limb darkening \citep{2000A&A...359..289C}. We adopted
the gravity-darkening law of \citet{2011A&A...533A..43E}. The
inclination angles follow a random distribution. The model suite is
limited to a minimum stellar mass of $1.7$ M$_\odot$ and it does not
account for the evolution of interacting binary systems.

Since the SYCLIST models are limited to high masses ($\geq 1.7$
M$_\odot$; suitable for covering the MSTO) and it is difficult to
constrain a cluster's binary fraction based on the morphology of its
MSTO, we exploited the less massive MS stars (see Fig.~\ref{fig:cmd})
to derive total binary fractions, $f_\mathrm{b}$, at masses where the
effects of stellar rotation are negligible. Binary models were
generated from the MIST isochrones by adding unresolved binaries with
different mass ratios, assuming a flat mass-ratio distribution. Next,
we compared the pseudo-colour distributions of stars in our synthetic
clusters with those in the observed clusters, using the same MS
selection criteria. The goodness of the comparison is given by the
Anderson--Darling $p$ value. We added a second parameter, $\sigma$, to
characterise the observational scatter in the pseudo-colour, combining
the effects of a possible internal age spread, photometric
uncertainties and differential extinction. A minor colour shift among
the clusters was also taken into consideration. We employed the
Markov-chain Monte Carlo method
\citep[\texttt{emcee};][]{2013PASP..125..306F} to determine the best
model and estimate the uncertainties in the resulting parameters.

We adopted this binary fraction as our input parameter and applied a
similar method to the MSTO stars to derive their rotation
distribution. In the SYCLIST models, on the zero-age MS (ZAMS) this
distribution is controlled by the ratio of the equatorial angular
velocity, $\Omega$, to the critical velocity
$\Omega_\mathrm{crit}$. The latter is the rotational angular velocity
where the surface gravity can no longer maintain equilibrium with the
centrifugal motion. We binned the resulting rotation rates into 10
bins, from $\Omega/\Omega_\mathrm{crit,ZAMS} = 0.0$ to 1.0 in steps of
0.1, and assigned 10 weights, $F_{\Omega_i}$. Each weight corresponded
to the fraction of stars with a given rotation rate.

We first generated a model cluster of 200,000 stars with a flat
rotation distribution. For each set of input parameters
($f_\mathrm{b}$, $F_{\Omega_i}$), we randomly selected the $N$ most
appropriate stars from the total pool to generate a new synthetic
cluster which satisfied the rotation distribution
required. Subsequently, we added scatter ($\sigma$) to the synthetic
cluster and selected the remaining MSTO stars for comparison with the
observed cluster. In practice, when selecting MSTO stars and
calculating their pseudo-colours, we shifted the ridge line towards
bluer colours by \unit[0.3]{mag} to include MSTO stars which have
scattered away from the MSTO region.

As for the MS region, a comparison of the pseudo-colour distributions
of MSTO stars in the synthetic and observed clusters yields the
corresponding $p_\mathrm{phot}$ values. We also used the rotational
velocities to address the degeneracy between the pseudo-colour and
rotation-velocity distributions (see Section~\ref{sec:rot}). For each
MSTO star with spectroscopic measurements, we selected the closest 100
stars in the synthetic cluster's CMD. We then computed the probability
of deriving the same $v\sin i$ distribution as observed. We first
resampled the measured projected rotational velocities according to
the error distribution and estimated the probabilities for this
alternative realisation. We iterated each run 100 times and adopted
the median value to minimise stochastic sampling effects. Finally, we
combined the corresponding $p_\mathrm{spec}$ values from the
spectroscopic data with $p_\mathrm{phot}$ and used \texttt{emcee} to
find the rotation model which best reproduces both the pseudo-colour
and the $v\sin i$ distributions. The uncertainties were derived from
the samples' 16th and 84th percentiles in the marginalised
distributions.

\subsection{Validation}

In this section, we examine the accuracy of our parameter recovery,
specifically for two crucial parameters, i.e., the binary fraction
($f_\mathrm{b}$) and the rotation rate ($F_{\Omega_i}$). We discuss
the importance of knowing the rotational velocity for the
determination of the rotation distribution.

\subsubsection{Binary fraction} \label{sec:binary}

\begin{figure}
\includegraphics[width=\columnwidth]{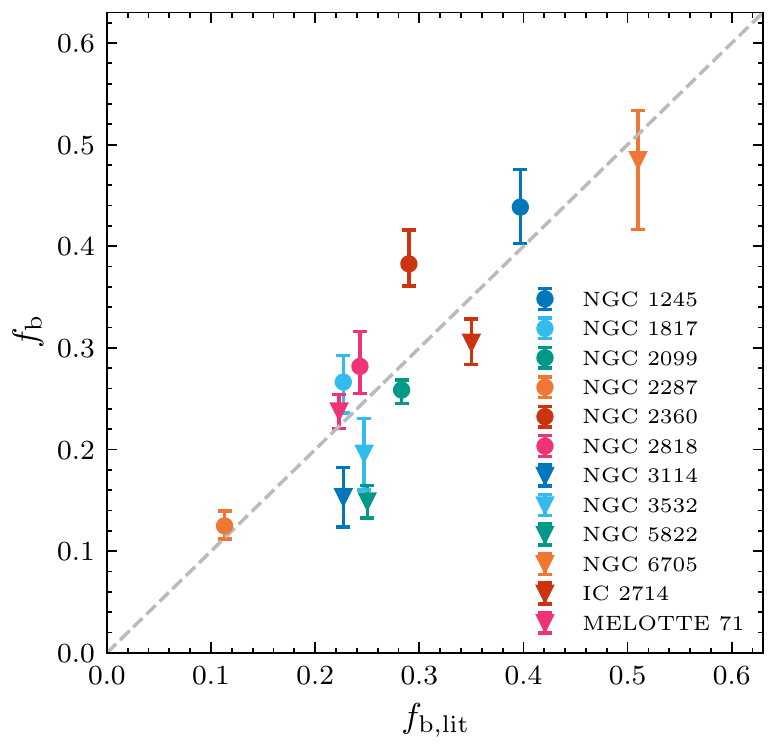}
\caption{Comparison of our newly derived binary fractions,
  $f_\mathrm{b}$, with the mean value of previously published
  results. The grey dashed line represents the linear, one-to-one
  relation. \label{fig:binary}}
\end{figure}

We applied our method to a few additional clusters to verify our
calculation of the binary fraction. We selected 12 OCs with binary
fraction measurements available in the literature
\citep{2005A&A...431..943B, 2018ApJ...869..139C, 2020MNRAS.491.2129D}
as our comparison sample. Given that these clusters are younger than
or of similar age as our sample OCs, our selection of MS stars is not
affected by any MSTO broadening and is thus suitable for validation
purposes. In Fig.~\ref{fig:binary}, we show our results (vertical
axis) versus the corresponding literature mean values (horizontal
axis). Considering possible differences in the sample selection and
the estimation method used, our estimates for these OCs are consistent
with the literature values

One possible flaw inherent to the derivation of binary fractions based
on \textit{Gaia} data is that some binaries could be (partially)
resolved, which is not taken into account in the application of this
method. If so, this would lead to an underestimation of the binary
fraction and this could be important for clusters which are
sufficiently close. However, this effect should not have a major
impact on our sample clusters. Even for the nearest cluster analysed
in this paper, IC 4756, only wide binaries with separations larger
than $\sim\unit[800]{AU}$ \citep[2 arcsec, adopted
  from][]{2018A&A...616A..17A} can be resolved by \textit{Gaia}, which
applies to less than 2 per cent of binaries in OCs
\citep{2020MNRAS.496.5176D}.

\subsubsection{Rotation rates} \label{sec:rot}

Verification of the rotation rates was done based on mock tests. We
generated mock data for various rotation distributions and applied our
method to verify whether the rotation rates were robustly
recovered. In particular, we generated a synthetic cluster with 200
MSTO stars (similar to the observed number) and a spectroscopic
completeness level of 25 per cent. The binary fraction and $\sigma$
were set at 20 per cent and \unit[0.02]{mag}, respectively.

\begin{figure*}
\includegraphics[width=0.8\textwidth]{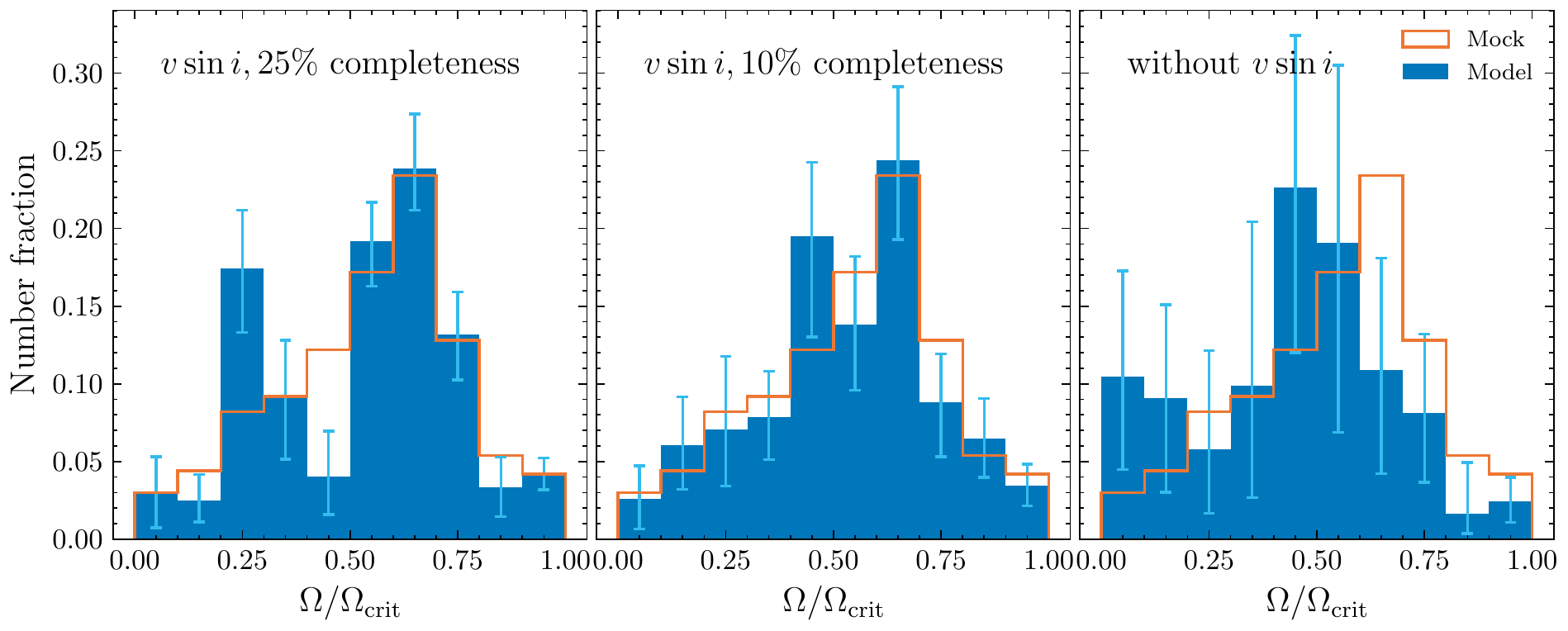}
\includegraphics[width=0.8\textwidth]{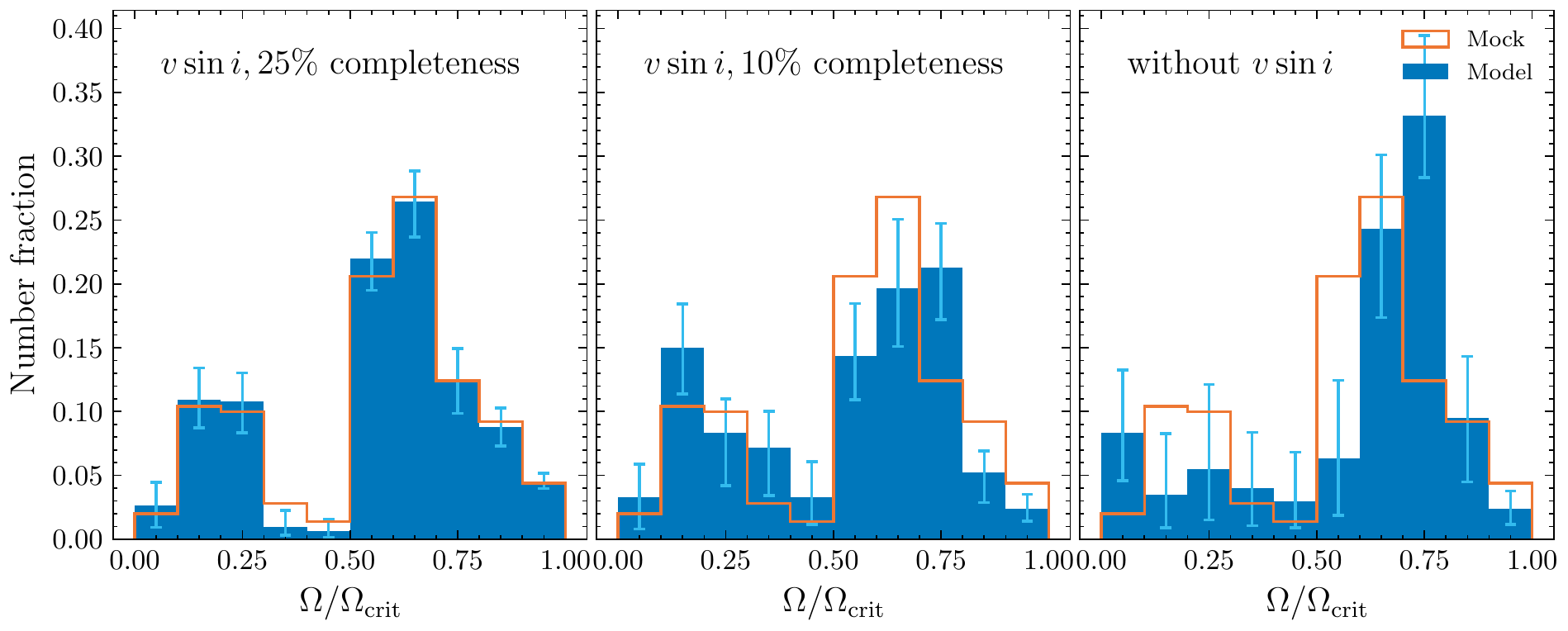}
\caption{Recovered rotation distributions from our mock tests. Two
  synthetic models of different rotation distributions are shown in
  the top and bottom rows. Histograms of the input mock data are
  compared with the best-fitting models (blue). The fits were done for
  different levels of $v\sin i$ completeness, including (left) 25 per
  cent, (middle) 10 per cent, and (right) 0. \label{fig:mock}}
\end{figure*}

Fig.~\ref{fig:mock} shows two representations of our mock tests. One
is based on a skewed normal distribution (first row) and the other is
characterised by a bimodal distribution (second row). The input and
recovered rotation rates are shown in the left-hand column as the
orange and blue histograms, respectively. Our derived rotation rates
are consistent with the mock data's true values (most fall within
$\unit[1]{\sigma}$).

We also checked the performance of our approach in the presence of
less or no spectroscopic information. The results for synthetic
clusters with 10 and 0 per cent of $v\sin i$ information are shown in
the middle and right-hand columns, respectively. The recovered
rotation rates exhibit strong deviations from the input
distribution. Although a general trend can tentatively still be
produced, the corresponding accuracy is far from
satisfactory. Although analyses of massive MC clusters solely based on
photometric data are practical \citep[e.g.,][]{2019ApJ...887..199G},
one should be careful when applying this approach to Galactic
OCs. This final test highlights the degeneracy resulting from fitting
models to poorly sampled data and underscores the need for additional
information (such as $v\sin i$ measurements).

\section{Results and Discussion} \label{sec:discussion}

\begin{figure*}
\centering
\sbox{\bigpicturebox}{\scalebox{1}[1]{\includegraphics[width=.66\textwidth]{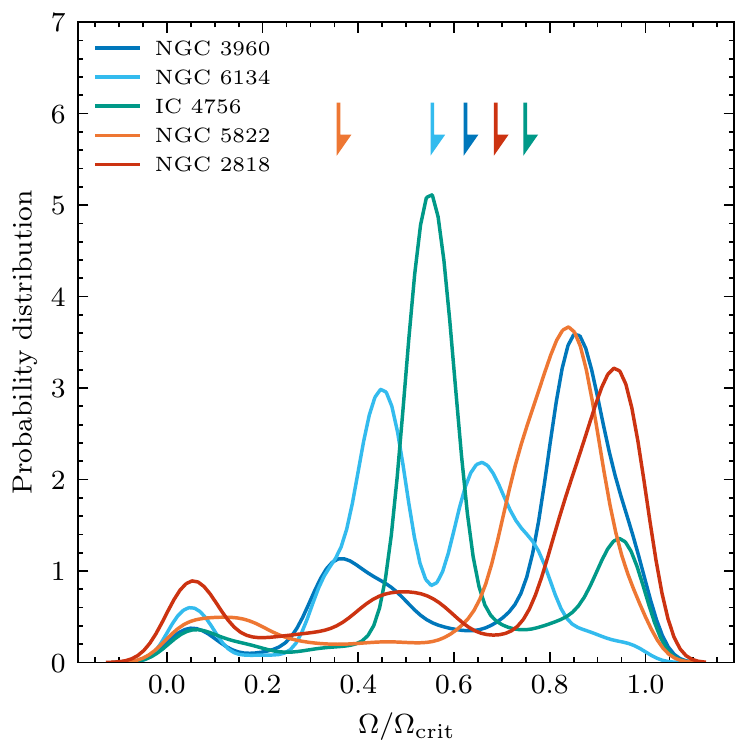}}}
\usebox{\bigpicturebox}
\begin{minipage}[b][\ht\bigpicturebox][s]{.32\textwidth}
\vspace{6pt}
\includegraphics[width=1\textwidth]{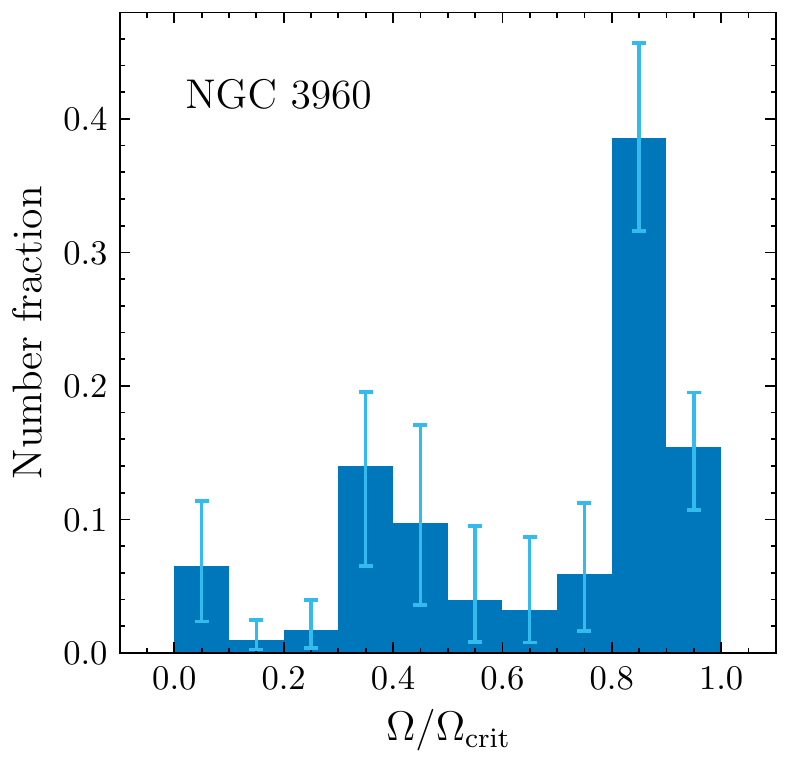}
\includegraphics[width=1\textwidth]{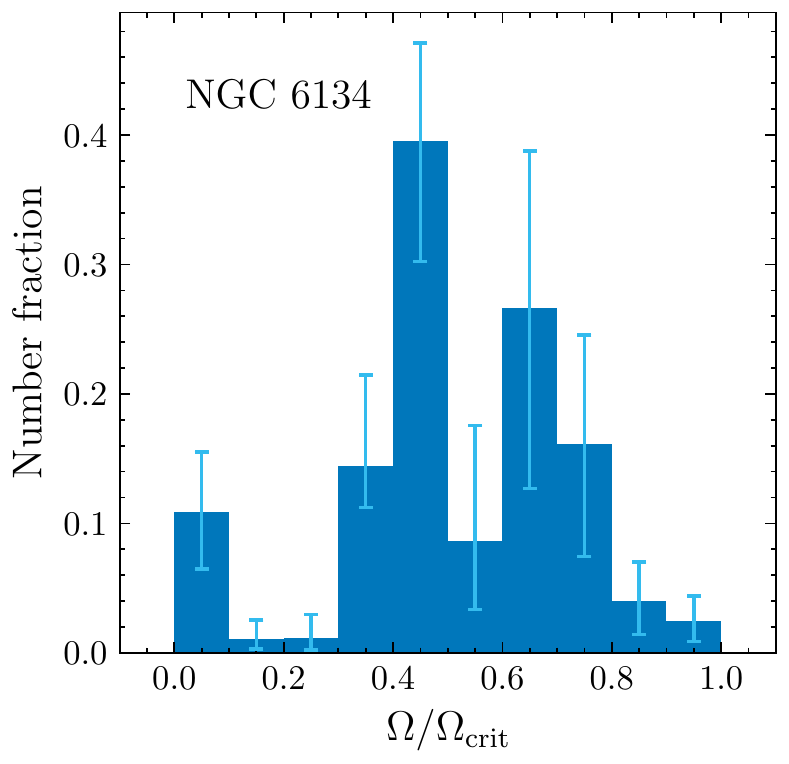}
\end{minipage}
\medskip
\includegraphics[width=.32\textwidth]{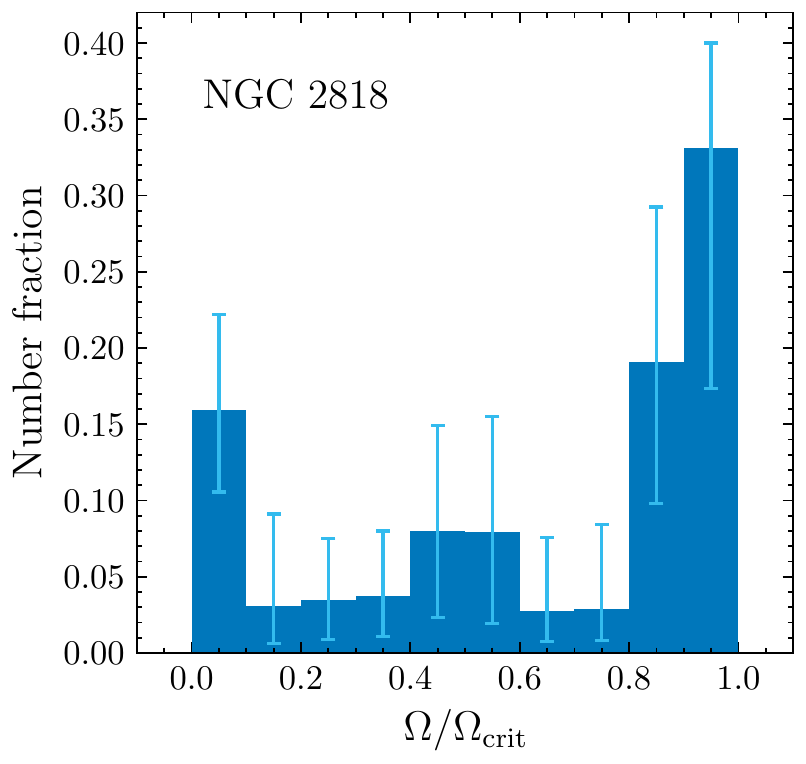}\hfill
\includegraphics[width=.32\textwidth]{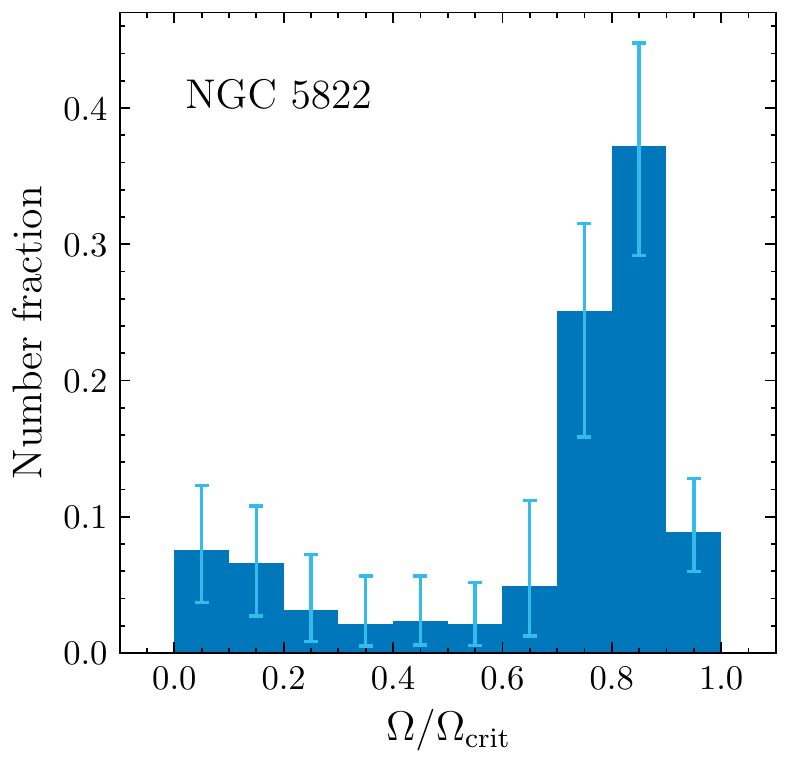}\hfill
\includegraphics[width=.32\textwidth]{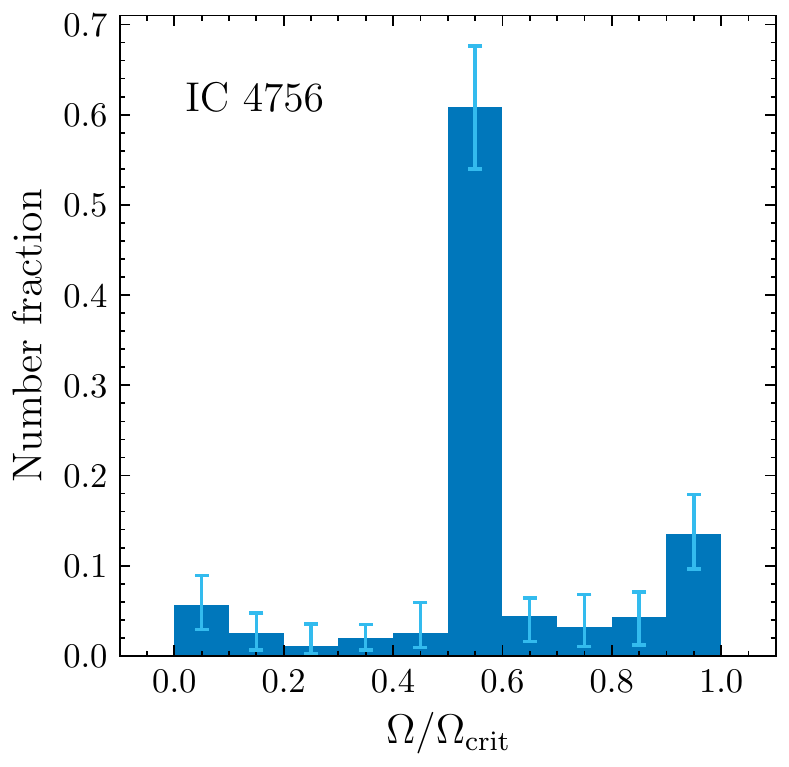}
\caption{(top left) Rotation rate ($\Omega_i/\Omega_\mathrm{crit,
    ZAMS}$) probability distribution of the MSTO stars according to
  the best-fitting models. The arrows indicate the velocities adopted
  for selection of the slowly and rapidly rotating subsamples in each
  of our clusters, corresponding to the local minima of the (mostly)
  bimodal distributions. (small panels) Rotation-rate distributions
  (bottom, top axes) for (clockwise from top right) NGC 3960, NGC
  6134, IC 4756, NGC 5822 and NGC 2818.\label{fig:rot_eq}}
\end{figure*}

The best-fitting rotation rates' weights are shown in
Fig.~\ref{fig:rot_eq} (small panels), whereas the rotation rate
probability distributions of the MSTO stars for all clusters are shown
in the top left-hand panel. All five clusters contain a significant
fraction of fast rotators. However, the locations and fractions of
fast rotators vary among the clusters. Whereas some clusters, like NGC
6134 and NGC 2818, appear to host three rotational populations
(including two slowly rotating populations), it is difficult to
reliably separate between populations of slow rotators because of
their minor colour differences. Therefore, we exercise caution and
suggest that these clusters may only contain two distinct populations
(i.e., a bimodal distribution) in stellar rotation space.

\begin{figure}
\includegraphics[width=\columnwidth]{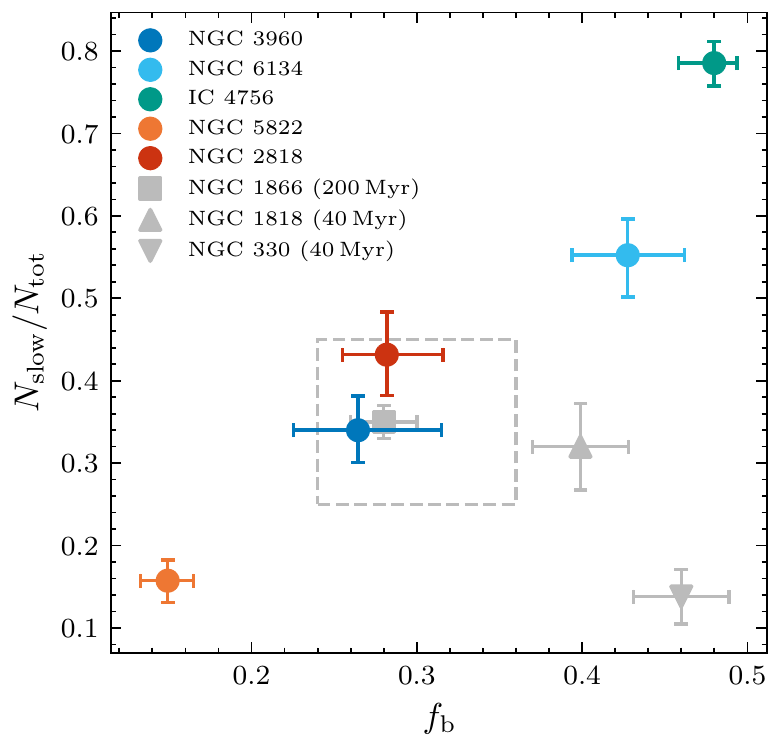}
\caption{Number fraction of slow rotators to the total number of MSTO
  stars in the best-fitting synthetic cluster versus their binary
  fractions. The uncertainties in $N_\mathrm{slow}/N_\mathrm{tot}$
  were estimated by resampling the best-fitting model parameters. The
  correlation between $N_\mathrm{slow}/N_\mathrm{tot}$ and
  $f_\mathrm{b}$ suggests a binary-driven mechanism behind the
  rotation rates. The majority of MC clusters reside in the dashed
  rectangle. Three young MC clusters---NGC 1866, NGC 1818 and NGC
  330---are shown using grey symbols (their ages are included in the
  legend). The deviations of these younger clusters from the
  correlation suggest that they may be at an earlier stage of tidal
  braking.
\label{fig:slow_ratio}}
\end{figure}

The number fraction of slow rotators,
$N_\mathrm{slow}/N_\mathrm{tot}$, among synthetic MSTO stars shows a
positive correlation with the clusters' total binary fraction
$f_\mathrm{b}$: see Fig.~\ref{fig:slow_ratio}. We verified that our
approach to selecting slowly and rapidly rotating subsamples, within
reasonable ranges, minimally affects this correlation by changing the
velocity adopted for our subsample selection to fixed values. In all
tests, the correlation between the number fraction of slow rotators
and the binary fractions of the clusters remains almost the same, {\it
  modulo} minor shifts in the absolute value of
$N_\mathrm{slow}/N_\mathrm{tot}$ In fact, one slow rotator
  ($v\sin i=\unit[123.1]{km\,s^{-1}}$) in NGC 6134 (\textit{Gaia} DR2
  ID 5941409684301615360) exhibited variation in its radial velocity
  during two observation epochs, spanning three days. The observed
  variation is around $\unit[80]{km\,s^{-1}}$, which is four times
  larger than the uncertainty associated with the velocity
  measurements. The estimated rotational velocities remained
  unchanged, with differences of less than the $v\sin i$ error
  ($\unit[7]{km\,s^{-1}}$). If this radial-velocity variation were
  induced by binaries, the star is likely a member of a close binary
  system with a separation of a few tens of solar radii. To arrive at
  this estimate, we assumed that its luminosity is not severely
  affected by the companion and we adopted the observed variation as
  the amplitude of radial velocity curve. Follow-up time-series
  observations are required to confirm this star's nature.

The approximately constant number ratios of bMS stars found in massive
MC clusters \citep{2018MNRAS.477.2640M} may result from the similar
binary fractions ($\sim 0.3$) which are prevalent in MC clusters
\citep{2009A&A...497..755M} if the correlation of
Fig.~\ref{fig:slow_ratio} also pertains to more massive clusters. For
instance, \citet{2017MNRAS.465.4363M} estimated that the number ratio
of bMS stars in NGC 1866 is 35 per cent; the cluster's overall binary
fraction is 0.28. These properties will place it comfortably on the
apparent correlation. However, some younger clusters, e.g., NGC 1818
($\sim\unit[40]{Myr}$) and NGC 330 ($\sim\unit[40]{Myr}$), do not
follow the correlation if we adopt their binary fractions from
\citet{2017ApJ...844..119L} and their number ratios from
\citet{2018MNRAS.477.2640M}. This may suggest that these young
clusters are still in the early stages of their stellar rotation
evolution, which may last for several tens of Myr. In NGC 1846
($\sim\unit[1.5]{Gyr}$), \citet{2020MNRAS.492.2177K} found that
$\sim45$ per cent of MSTO stars are slow rotators. They derived a
binary fraction of $\sim$7 per cent based on radial velocity
variations from four-epoch spectra. However, because their binary
fraction measurement is limited by their temporal coverage and most
sensitive to the detection of tight binaries, a large fraction of the
cluster's binaries are likely missing. Therefore, we did not
include this cluster in Fig.~\ref{fig:slow_ratio}.

To date, none of the prevailing theoretical models
\citep[e.g.,][]{2015MNRAS.453.2637D, 2020MNRAS.495.1978B} can
naturally explain the observed correlation. Our observational result
is in apparent conflict with the model of \citet{2020MNRAS.495.1978B},
which suggested that binaries may destroy stellar discs, thus
resulting in faster rotational velocities on the MS. Moreover, this
interpretation cannot explain the cluster-to-cluster variations in the
MSTO pseudo-colour distributions in Fig.~\ref{fig:cmd} (top right),
which we verified using a two-sample Kolmogorov--Smirnov test. This is
so, because our result is not an artefact of requiring a higher binary
fraction to compensate for the potential narrowing of the MSTO owing
to a lower fraction of fast rotators.

Neither does our correlation follow the scenario of
\citet{2015MNRAS.453.2637D}, where all or most stars form as fast
rotators with a fraction undergoing close-binary tidal
braking. Although this model predicts a larger fraction of slow
rotators for a higher binary fraction, everything else being equal,
the fraction of slow rotators can never exceed the binary
fraction. Since only close binaries may become tidally locked, only a
small subset of binaries could contribute. However, as shown in
Fig.~\ref{fig:slow_ratio}, the slow rotators' number ratios are
comparable to or even higher than the binary fractions. Moreover, the
slope between $N_\mathrm{slow}/N_\mathrm{tot}$ and $f_\mathrm{b}$ is
close to unity regardless of the subsample selection, thus suggesting
an alternative binary-driven mechanism. However, we cannot rule
  out a possible origin associated with the initial phase. The slow
  rotators we observe at the present time originate from both the
  initial population of slow rotators and may also have evolved from
  the fast rotators. \citet[][their figure 7]{2010ApJ...722..605H}
  selected a young subpopulation of B-type stars that just evolved
  from the zero-age main sequence and showed that a large fraction of
  them were formed as slow rotators. So far, it is unclear which
physical property of (close and wide) binaries determines the rotation
distributions. This may explain why \citet{2020MNRAS.492.2177K} did
not find any significant differences among the binary fractions of
slow and fast rotators. If we consider the possibility of missing
binaries from their sample, combined with the disruption of wide
binaries, there could be distinct differences between slow and fast
rotating populations.

If binaries are the dominant drivers of slow rotators, one would
expect slow rotators to be more centrally concentrated in a cluster
because they are generally more massive than their rapidly rotating
counterparts, and they would thus sink more easily to the cluster
centre owing to two-body relaxation \citep[dynamical mass
  segregation;][]{1987gady.book.....B}. However, this scenario appears
at odds with recent observations in young MC
clusters. \citet{2017ApJ...846L...1D} and \citet{2017MNRAS.465.4363M}
found that in NGC 1866 ($\sim\unit[400]{Myr}$), bMS stars (slow
rotators) are less centrally concentrated than rMS stars (fast
rotators). Meanwhile, in NGC 1856 ($\sim\unit[300]{Myr}$-old), the
number ratios of bMS and rMS stars remain unchanged at different
radii, thus suggesting that they are spatially homogeneously
distributed \citep{2017ApJ...834..156L}.

We used the best-fitting synthetic cluster to infer the rotational
velocities for all member stars \citep{2019ApJ...876..113S} and
employed the spatial locations of the observed stars to estimate their
degree of mass segregation. Because of their low number densities, it
is hard to robustly determine the centres of OCs. Therefore, we
adopted minimum spanning trees (MSTs) to quantify a cluster's degree
of mass segregation \citep{2009MNRAS.395.1449A}. The mass segregation
ratio, $\Lambda_\mathrm{MSR}$, of a given population is defined as the
ratio of the average random path length, $l_\mathrm{random}$, to that
of the entire population, $l_\mathrm{pop}$,
\begin{equation}
\Lambda_\mathrm{MSR} = \frac{\langle l_\mathrm{random} \rangle}{l_\mathrm{pop}}
\pm \frac{\sigma_\mathrm{random}}{l_\mathrm{pop}},
\end{equation}
where $l$ represents the length of the shortest path connecting all
data points and $\langle l_\mathrm{random} \rangle\pm
\sigma_\mathrm{random}$ quantifies the length distribution. We
determined the OCs' MSTs using \texttt{MiSTree}
\citep{2019JOSS....4.1721N}, in celestial coordinates. Given our OCs'
close proximity and the low stellar densities, the impact of sampling
incompleteness of the low-mass stars in estimating
$\Lambda_\mathrm{MSR}$ is negligible.

To link our results with the observations of blue and red MSs, we
  adopted bMS stars as those stars with $v\sin i <
  \unit[150]{km\,s^{-1}}$, whereas rMS stars have $v\sin i >
  \unit[150]{km\,s^{-1}}$. This velocity threshold was selected based
  on the velocity dip at $v\sin i\approx \unit[100]{km\,s^{-1}}$
  observed in NGC 1846 \citep{2020MNRAS.492.2177K}. Given the mass
  differences of MSTO stars in clusters with different ages, this
  corresponds to $v\sin i\approx \unit[150]{km\,s^{-1}}$ for our
  sample. We assumed that the rotation rate
  ($\Omega/\Omega_\mathrm{crit}$) of the dip does not change for
  clusters of either \unit[1]{Gyr} or \unit[1.5]{Gyr} (NGC 1846). The
  critical rotation velocity only marginally decreases as a star
  evolves, whereas the rotation velocity strongly depends on stellar
  mass \citep[][their figure 1]{2020MNRAS.495.1978B}. The typical
MSTO stellar mass is around $\unit[1.5-1.6]{\mathrm{M}_\odot}$ and
$\unit[1.8]{\mathrm{M}_\odot}$ for NGC 1846 and our cluster sample,
respectively. Based on \citet{2014A&A...566A..21G}, the velocity dip
for bMS and rMS should be $\sim$1.5 times larger in a 1 Gyr-old
cluster than in NGC 1846.

The corresponding $\Lambda_\mathrm{MSR}$ values are listed in
Table~\ref{tab:oc}. In all clusters, except for NGC 2818, the bMS
stars exhibit significant spatial segregation, $\Lambda_\mathrm{MSR} >
1$. Moreover, $\Lambda_\mathrm{MSR, b} > \Lambda_\mathrm{MSR, r}$ in
NGC 3860 and NGC 5822, with differences $> 2\sigma$. This suggests
that the bMS stars in these clusters are more centrally concentrated
than their rMS counterparts. This, hence, offers promising supporting
evidence of a more massive origin of the bMS stars in these
OCs. However, this is at odds with MC cluster results
\citep[e.g.,][]{2017MNRAS.465.4363M, 2020MNRAS.492.2177K}. The
difference could be attributed to binary disruption in the cluster
centre. \citet{2017MNRAS.465.4363M} found the binary fraction to
increase towards the cluster outskirts, following the same radial
distribution as the bMS stars in NGC 1866, whereas this effect is not
obvious in OCs.

In summary, we have discovered a strong observational correlation
between the number ratios of the slow rotators and the binary
fractions in five Galactic OCs and shown support for a marked
concentration of bMS stars. Future work will extend the survey to
younger clusters, cover a larger parameter space (e.g., in
$f_\mathrm{b}$, stellar and cluster mass, metallicity, etc.) and study
the detailed history of stellar rotation in cluster environments to
gain additional important insights into the star cluster formation
processes.

\section*{Acknowledgements}
\addcontentsline{toc}{section}{Acknowledgements}

W.S. thanks Xiao-Wei Duan for insightful discussions. He is grateful
for financial support from the China Scholarship
Council. L.D. acknowledges research support from the National Natural
Science Foundation of China through grants 11633005, 11473037 and
U1631102. This work has made use of data from the European Space
Agency (ESA) mission \textit{Gaia}
(\url{http://www.cosmos.esa.int/gaia}), processed by the \textit{Gaia}
Data Processing and Analysis Consortium (DPAC,
\url{http://www.cosmos.esa.int/web/gaia/dpac/consortium}).


\section*{Data Availability}

The data analysed in this paper will be shared upon request.







\appendix
\section{Dynamical timescale} \label{sec:dynamical}

Dynamical modelling of OCs is rendered uncertain by the small numbers
of their member stars. To derive approximate dynamical ages for our
sample OCs, we calculated their half-mass relaxation time-scales
\citep{1987A&A...184..144M},
\begin{equation}
t_\mathrm{rh} = 8.92\times10^5\frac{M_\mathrm{tot}^{1/2}}{\bar{m}}\frac{r^{3/2}}{\ln(0.4 M_\mathrm{tot}/\bar{m})} \unit{yr},
\end{equation}
where $r$ is the half-mass radius derived from number counts,
$M_\mathrm{tot}$ is the total mass derived following
\citet{2019ApJ...876..113S} and $\bar{m}$ is the typical mass of stars
in the cluster. This estimate yields relatively similar half-mass
relaxation time-scales for all of our sample OCs, ranging from
$\sim\unit[60]{Myr}$ to $\sim\unit[120]{Myr}$. We double checked this
result by comparison with \citet{1987gady.book.....B}
\begin{equation}
t_\mathrm{relax} = \frac{N}{8 \ln N} t_\mathrm{cross},
\end{equation}
where $t_\mathrm{cross}=r/\sigma_v$ is the crossing time, $N$ is the
total number of stars and ${\sigma_v}$ is the velocity dispersion. We
adopted the latter from \citet{2018A&A...619A.155S}. Both estimates
are in reasonable mutual agreement, in the sense that the clusters'
dynamical time-scales are around \unit[100]{Myr}, varying by a factor
of up to 2. This means that our clusters have evolved through
approximately 9--16 half-mass relaxation time-scales and share similar
dynamical ages.


\bsp 
\label{lastpage}
\end{document}